\documentclass{article}
\usepackage{graphicx} 
\usepackage{svg}

\usepackage{amsmath}
\usepackage{parskip}
\usepackage{float}
\usepackage[a4paper, total={5.5in, 8in}]{geometry}
\usepackage{authblk}

\usepackage{tabularx}

\usepackage{cleveref}

\usepackage[
    backend=biber,
    style=authoryear,
    maxcitenames=2,      
    maxbibnames=99,      
    uniquelist=false,
    giveninits=true,     
    terseinits=true,     
    dashed=false,        
    isbn=false,          
    doi=true,            
    uniquename=false,
    url = false,
    eprint = false]{biblatex}
\title{Generalized modeling reveals stabilizing mechanisms in cross-feeding microbial communities}
\author[a,b]{Tom Clegg}
\author[a,b,c]{Thilo Gross}
\date{}

\affil[a]{Helmholtz Institute for Functional Marine Biodiversity at the University of Oldenburg, Im Technologiepark 5, 26129 Oldenburg, Germany.}
\affil[b]{Alfred Wegner institute, Am Handelshafen 12, 27570 Bremerhaven, Germany}
\affil[c]{Institute for Chemistry and Biology of the Marine Environment, Carl-von-Ossietzky-Straße 9-11, 26111 Oldenburg, Germany}

\begin{document}

\maketitle
\section*{Abstract}
Understanding what makes microbial communities stable and able to resist perturbation is a fundamental question in microbial ecology. Theoretical approaches to this problem typically require making strict assumptions about the structure, parameters and functional forms used in models, hindering general insight. Here, we address this problem by developing and analyzing a generalized model of microbial community dynamics, capturing resource-mediated competitive and cross-feeding interactions between populations. The generalized modeling approach allows us to assess the stability of microbial systems across large ensembles of models with variable size, structure and non-linear dynamics. Our analysis shows how community stability is constrained by system size and identifies its key metabolic and dynamical drivers, highlighting the central roles of self-limitation, metabolic efficiency and plasticity. Furthermore, we show how the structure of communities mediates their stability, allowing communities to counteract the destabilizing cross-feeding loops through organization into hierarchical trophic structures. Together, these results provide new insight into the stability of microbial communities and offer a flexible analytical framework for further work to explore the responses of communities to perturbation. 

\clearpage

\section*{Introduction}
Understanding the mechanisms that govern community stability is a fundamental challenge in ecology \parencite{Allesina2015}. At the core of this topic is May's seminal work, demonstrating that large, random communities tend to be less stable as they increase in size and complexity \parencite{may_will_1972}. Although this prediction appears to be odds with the diversity ecosystems we see in nature, this contradiction can be resolved by recognizing that specific non-random structural and dynamical features must underpin community stability \parencite{May2001, coyte_ecology_2015}. This insight set the stage for a large body of research over the subsequent decades, aiming to identify these stabilizing factors across various ecological contexts, including food webs \parencite{gross_generalized_2009, Allesina2012}, mutualistic networks \parencite{thebault_stability_2010} and more recently, microbial communities \parencite{coyte_ecology_2015, butler_stability_2018}.

Identifying general stabilizing features in microbial communities poses a significant challenge due to their unique resource-mediated dynamics. Microbes interact via a shared chemical environment, competing for substrates while simultaneously engaging in facilitative cross-feeding, where one species' by-products serve as another's resources \parencite{dsouza_ecology_2018}. Furthermore, microbes display high plasticity, switching metabolic pathways and rewiring their interaction networks in response to environmental fluctuations \parencite{monod_growth_1949,solopova_bet-hedging_2014,pacciani-mori_dynamic_2020}. Together these features mean that microbial community dynamics are both complex and highly variable, making it hard to apply results from standard theoretical ecology.

Recent work has tackled this problem by extending explicit consumer-resource frameworks to model microbial communities. Building on classical ecological theory, this work has successfully demonstrated how specific structural and dynamical features can enhance can microbial persistence and stability \parencite{butler_stability_2018, marsland_available_2019, goldford_emergent_2018, mentges_microbial_2020}. For example, studies have shown how community stability conditions can be derived under assumptions of structural symmetry in consumer-resource interactions \parencite{gibbs_stability_2022}, and how dynamic metabolic pathway-switching fosters species coexistence \parencite{pacciani-mori_dynamic_2020}. While these explicit models have provided crucial insight, their conclusions often rely on assuming specific functional forms (e.g., Monod or Michaelis-Menten kinetics), parameter values or fixed community structures. As such, it is hard to find general principles of stability independent of our specific modeling choices.

Generalized modeling offers a direct solution to this problem. It provides a mathematical framework for evaluating the stability of complex non-linear systems without requiring explicit kinetic equations \parencite{gross_generalized_2006, massing_generalized_2022}. By normalizing system dynamics in terms of qualitative \textit{generalized parameters} with clear ecological interpretation, this framework allows for the simultaneous analysis of entire ensembles of potential models under a given set of structural constraints. Having successfully identified universal stabilizing mechanisms in complex food webs \parencite{gross_generalized_2009, anderson_body_2021} and motifs of microbial interactions \parencite{massing_generalized_2026}, generalized modeling provides an ideal approach for finding general stabilizing principles in resource-mediated microbial systems.

In this paper we apply the generalized modeling framework to investigate microbial community dynamics and explore how the structural and dynamical features of these systems determine their stability. We first develop a generalized model of microbial cross-feeding dynamics based on previously used consumer-resource models. Then, we demonstrate how stability scales with the size of the community and ratio of consumers to resources, showing how the probability of stability falls as the number of consumers increases. We then explore how further dynamical and structural features of the community influence stability, highlighting the specific mechanisms that stabilize communities within the space of feasible system sizes. Finally we demonstrate how community structure, specifically the organization into hierarchical trophic structures can increase stability by breaking destabilizing feedback loops caused by cross-feeding between populations.

\section*{Methods}
\subsection*{The Microbial Community Model}
We develop a generalized model of a microbial community, representing their resource-mediated competition and cross-feeding dynamics. The model structure is based on widely-used microbial consumer-resource models \parencite{goldford_emergent_2018,marsland_available_2019,mentges_microbial_2020}.
\begin{align}
  \dot{X}_i &= G_i(X_i, \mathbf{Y}) - L_i(G_i) - M_i(X_i) \label{eq:dX}, \\
  \dot{Y}_a &= I_a(Y_a) - O_a(Y_a) - \sum_i U_{ia}(X_i, \mathbf{Y}) + \sum_i K_{ia}(X_i, \mathbf{Y}). \label{eq:dY}
\end{align}
The equations consist of a set of non-linear functions representing the flux of material between $N$ consumers, $X_i$, and $M$ environmental resource pools, $Y_a$, where $\mathbf{Y} = (Y_1,\dots,Y_M)$ denotes the vector of resources. Consumers grow by taking up resources at rate $G_i$, which is a function of resource availability and the biomass of the focal population. Consumers release some of these resources back into the environment as by-products at rate $L_i$ and experience an additional mortality $M_i$. Each resource is externally supplied to the system at rate $I_a$ and removed from the system at rate $O_a$, representing outflow or degradation. Resources are taken up from the environment by each consumer at rate $U_{ia}$, which represents the uptake of resource $a$ by consumer $i$, and released back into the environment as by-products at rate $K_{ia}$, representing the flux of resource $a$ generated by consumer $i$. 

\Cref{eq:dX,eq:dY} capture the general structure of microbial community dynamics but require additional constraints to represent the ecology of these systems. The first is the link between consumer uptake and by-product release. We assume by-products are created from the metabolic activity in the cell, the rate of which is governed by total substrate uptake rate $G_i$. As such, we introduce the quantity $\Lambda_i$, which is the total proportion of uptake that is converted into by-products, sometimes referred to as leakage \parencite{goldford_emergent_2018}, and is defined as
\begin{align}
    \Lambda_i(G_i) = \frac{L_i(G_i)}{G_i(X_i, \mathbf{Y})},
\end{align}
allowing the rate of by-product release to be expressed directly as a proportion of the overall uptake rate $L_i = \Lambda_i G_i $. In contrast to previous implementations of microbial consumer resource models \parencite{goldford_emergent_2018,marsland_minimal_2020}, we express this leakage term not as a constant, but as a function of the uptake rate $G_i$. This lets us consider regimes where consumer efficiency varies in response to resource availability, for example, in the case of overflow metabolism where microbes can become less efficient when resources are abundant \parencite{basan_overflow_2015}.

The second is the link between consumer uptake gains $G_i$ and the associated resource pool fluxes $U_{ia}$ and $K_{ia}$. To quantify this relationship, we introduce the auxiliary variable $S_i$, which represents the total amount of substrate in the environment available to consumer $i$. We define $S_i$ as the sum
\begin{align}
    S_i(\mathbf{Y}) = \sum_a F_{ia}(Y_a),
\end{align}
where $F_{ia}(Y_a)$ is the contribution of resource $a$ to the total amount of resources available to population $i$. $F_{ia}(Y_a)$ captures the influence of both the amount of a given substrate in the environment and the resource preferences of the consumer. Because $F_{ia}(Y_a)$ is expressed as a function of resource availability, it represents metabolic plasticity, by varying resource utilization based on the environment. Assuming that resource availability affects growth solely through this aggregate substrate term, we can also write $G_i(X_i, \mathbf{Y})=G_i(X_i,S_i)$.

Given that the consumer uptake gain from resource $a$ is equal to the loss from that resource's environmental pool, we can express $U_{ia}$ as the proportion of the total uptake $G_i$ made up by resource $a$
\begin{align}
    U_{ia}(X_i, \mathbf{Y}) = \frac{F_{ia}(Y_a)}{S_i(\mathbf{Y})} G_i(X_i, S_i).
\end{align}

We can also apply the same reasoning to the resource leakage term $K_{ia}$. Following \textcite{marsland_available_2019}, we define a global stoichiometric matrix $D_{ab}$ that captures the fraction of resource $a$ converted into by-product $b$ upon consumption, such that $\sum_b D_{ab} = 1$. Together with the consumer inefficiency term $\Lambda_i$, this allows us to write $K_{ia}$ as the total flux of by-product $a$ that is produced by consumption across all other resources (captured in the sum over $b$)
\begin{align}
    K_{ia}(X_i,\mathbf{Y}) &= \sum_b U_{ib}(X_i, \mathbf{Y}) \Lambda_i(G_i) D_{ba}, \nonumber \\
    &= \sum_b \frac{F_{ib}(Y_b)}{S_i(\mathbf{Y})} G_i(X_i, S_i)  \Lambda_i(G_i) D_{ba}.
\end{align}

\subsection*{Generalized Modeling}
With the structure of the model and the additional ecological constraints defined, we analyze the stability properties of the microbial community model using the generalized modeling framework \parencite{gross_generalized_2006,massing_generalized_2022}. This approach allows us to assess stability by computing the Jacobian matrix for the system described in \cref{eq:dX,eq:dY} without having to make specific assumptions about the functional forms used. A full detailed derivation can be found in the supplementary material.

The Generalized modeling procedure starts with the normalization of the dynamic equations relative to an equilibrium point. Assuming that a fixed point exists, we define the normalized state variables $x_i = X_i / X_i^*$ and $y_i = Y_a / Y_a^*$, using lowercase symbols for normalized quantities and the asterisk ($^*$) for equilibrium quantities. We also define the corresponding normalized flux functions so that, for example, the normalized consumer uptake function is $g_i(x_i,s_i) = G_i(X_i,S_i) / G_i(X_i^*,S_i^*) $. Substituting these into \cref{eq:dX,eq:dY} yields the full set of normalized equations for community dynamics 
\begin{align}
    \dot{x}_i &= \alpha_i \biggl[ g_i(x_i,s_i) - \Lambda_i^* g_i(x_i,s_i) \lambda_i(g_i) - (1- \Lambda_i^*) m_i(x_i) \biggr] \label{eq:norm_x}, \\
    \dot{y}_i &= \beta_a \biggl[\rho_a i_a(y_a)
    - \sigma_a o_a(y_a) 
    - \tilde{\sigma}_a \sum_i \gamma_{ia} u_{ia}(x_i, \mathbf{y})
    + \tilde{\rho}_a \sum_i \eta_{ia} k_{ia}(x_i, \mathbf{y}) \biggr],\label{eq:norm_y}
\end{align}
where the normalized loss ($u_{ia}$) and gain ($k_{ia}$) of resources due to uptake and by-product release can again be expressed in terms of consumer uptake as
\begin{align}
    u_{ia}(x_i, \mathbf{y}) &= \frac{f_{ia}(y_a)}{s_i(\mathbf{y})} g_i(x_i,s_i), \\
    k_{ia}(x_i, \mathbf{y}) &=  \sum_b   \frac{f_{ib}(y_b)}{s_i(\mathbf{y})} g_i(x_i,s_i) \lambda_i(g_i) \phi_{iba}.
\end{align}
The normalized equations describe the dynamics of the system around the fixed point and are governed by the normalized functions as well as a new set of terms called \textit{scale} and \textit{branching} parameters (\Cref{tab:params}). 

The scale parameters $\alpha_i$ and $\beta_a$ capture the characteristic timescales of the consumer population and resource pools, respectively. These quantify the turnover rate of each system component at steady state.

The branching parameters represent the relative sizes of the fluxes between the consumer and resource components. These capture the structure of the network of resource uptake and release in the system. For consumers, the leakage parameter $\Lambda_i^*$ dictates the total inefficiency of consumer $i$ at equilibrium, the fraction of total uptake released as by-products. 

In the resource equations, the parameters $\rho_a$ and $\tilde{\rho}_a$ partition resource influx between external supply and metabolic by-product release, while $\sigma_a $ and $\tilde{\sigma}_a$ represent the relative contributions of outflow and consumer uptake to total resource loss. These equations also include the branching parameters $\gamma_{ia}$ and $\eta_{ia}$, which encode the network structure of the individual consumer-resource interactions by representing the proportion of the total uptake and by-product release of resource $a$ performed by consumer $i$, respectively. Finally, the parameter $\phi_{iba}$ quantifies the fractional contribution of the consumption of resource $b$ to the total production of by-product $a$ by consumer $i$, capturing the structure of the joint uptake and by-product release matrices.

\begin{table}[H]
\renewcommand{\arraystretch}{1.2}
\caption{Generalized parameters and their ecological interpretations.}
\begin{tabularx}{\textwidth}{lX}
\hline
\textbf{Parameter} & \textbf{Interpretation} \\ \hline
\multicolumn{2}{l}{\textit{Scale Parameters (Relative Timescales)}} \\
$\alpha_i$ & Relative timescale of the population dynamics for consumer $i$ \\
$\beta_a$ & Relative timescale of the dynamics for resource $a$ \\ \hline
\multicolumn{2}{l}{\textit{Branching Parameters}} \\
$\Lambda_i^*$ & Leakage parameter; inefficiency of consumer $i$ at equilibrium \\
$\rho_a$ & Proportion of resource $a$ gain from the external environment \\
$\tilde{\rho}_a$ & Proportion of resource $a$ gain from by-product release \\
$\sigma_a$ & Proportion of resource $a$ loss due to system outflow \\
$\tilde{\sigma}_a$ & Proportion of resource $a$ loss due to consumer uptake \\ 
$\gamma_{ia}$ & Proportion of total uptake of resource $a$ performed by consumer $i$ \\
$\eta_{ia}$ & Proportion of total by-product release of resource $a$ performed by consumer $i$ \\
$\phi_{iba}$ & Proportion of consumer $i$'s production of resource $a$ generated from consumption of resource $b$ \\ 
\hline
\multicolumn{2}{l}{\textit{Elasticity Parameters}} \\
$g_i^{(x)}$ & Sensitivity of consumer uptake to biomass \\
$g_i^{(s)}$ & Sensitivity of consumer uptake to substrate availability \\
$m_i^{(x)}$ & Sensitivity of consumer mortality \\
$f_{ia}^{(y)}$ & Sensitivity of consumer resource plasticity \\
$\lambda_i^{(g)}$ & Sensitivity of consumer inefficiency \\
$i_a^{(y)}$ & Sensitivity of external resource supply \\
$o_a^{(y)}$ & Sensitivity of resource outflow and degradation \\
\hline
\label{tab:params}
\end{tabularx}
\end{table}

\subsection*{Jacobian Matrix}
 We compute the Jacobian matrix $\mathbf{J}$ from \cref{eq:norm_x,eq:norm_y} by linearizing the system around the equilibrium point. Because the normalized functions are defined relative to their equilibrium value, they are equal to 1 at this point (e.g., $g_i(x_i^*,s_i^*)= g_i(1,1) = 1$). This means they drop out of the equations, resulting in a Jacobian matrix with entries that are independent of the state variables, allowing us to avoid solving the equations explicitly for the steady-state. We show the full expressions for the Jacobian in the supplementary material.

By linearising the system, we obtain the final set of parameters of the generalized model: the elasticity parameters (\Cref{tab:params}). These are defined as the derivatives of the normalized functions at the equilibrium point and are supplied as parameters to the model, capturing the nonlinearity of processes central to the ecology of the microbial community. For example, the saturation of consumer uptake is captured in the sensitivity of uptake to substrate availability
\begin{align}
    g_i^{(s)} = \left. \frac{\partial g_i(x_i, s_i)}{\partial s_i}\right\rvert_*.
\end{align}
Here, the notation $g_i^{(s)}$ denotes the partial derivative of the normalized uptake function with respect to substrate availability, evaluated at the fixed point. An elasticity of 1 corresponds to a linear response around the fixed point. Values greater than 1 represent a ``supra-linear'' response, while values between 0 and 1 indicate a ``sub-linear'' or saturated response. In this example, the elasticity approaches a value of $0$ as consumer uptake becomes fully saturated.

For most functions, the elasticities are straightforward, representing the sensitivity of gain or loss terms to the abundance of consumers or resources. However, the elasticity of by-product release, $\lambda^{(g)}$, and the contribution of resources, $f^{(y)}$, are of particular interest as they represent the metabolic flexibility of consumers. $\lambda^{(g)}$ governs the sensitivity of resource-use inefficiency to consumer uptake, allowing us to represent processes such as overflow metabolism as discussed above. $f^{(y)}$ controls how much consumers alter substrate usage in response to changing availability. A value of $f^{(y)} = 1$ corresponds to a passive uptake regime, where consumers use resources proportional to their abundance (i.e., mass action). When $f^{(y)} > 1$, this represents active switching, in which consumers alter their uptake profiles to favor the most abundant resources (i.e., metabolic plasticity)

\subsection*{Generating Communities}
To explore the stability properties of microbial communities, we generate ensembles of model systems for which we can compute the Jacobian using the approach described above. Generating an individual realization of a community begins with defining the underlying structure of the resource consumption and exchange network. This structure, encoded in the combination of the uptake preference $\mathbf{A_U}$ and stoichiometric matrix $\mathbf{D}$, can be constructed with a generative model to incorporate specific structural features. We then supply the scale parameters $\alpha$ and $\beta$, as well as the total consumer uptake fluxes $G^*$ and efficiencies $\Lambda^*$, drawn from biologically plausible distributions. From these parameters and the matrices $A_U$ and $D$, we calculate the remaining branching parameters in \cref{eq:norm_x,eq:norm_y}, taking care that the parameters correspond to a feasible fixed point. A more technical explanation of this is provided in the supplementary material. We sample the elasticity parameters from distributions tailored to the specific ecological scenario being considered. Once the Jacobian matrix is computed, we assess stability by looking at the real part of the dominant eigenvalue, $\text{Re}(\lambda_{\text{max}})$. 

\section*{Results}
\subsection*{Consumer and resource diversity limit community stability}
To establish the broad-scale behavior of the model, we first examine how community size, defined by the number of consumers $N$ and resources $M$, influences stability. We generate an ensemble of communities with $N$ and $M$ sampled across 30 values the range $[20,200]$, with $500$ realizations for each $N,M$ pair, totaling $4.5 \times 10^5$ replicates. For each realization we generate uptake and by-product release networks with random topology and connectance. The scale and elasticity parameters are sampled uniformly from biologically feasible ranges (see Supplementary Table~1). 

In this initial analysis, we set the consumer self-limitation terms equal, $g^{(x)} = m^{(x)}$ and consider constant consumer efficiency $\lambda^{(g)}=0$. This places consumer populations at the limit of marginal stability, allowing us to isolate the structural effects of system size from the stabilizing effects of self-limitation. This choice is analogous to the structure of most consumer-resource frameworks \parencite{macarthur_species_1969,marsland_minimum_2020}, where uptake and mortality are linear with respect to consumer abundance (i.e, $g^{(x)} = m^{(x)} = 1$). We relax this assumption when we consider the general effects of the parameter variation in the subsequent section. 

\begin{figure}[H]
    \centering
    \includegraphics[width = \textwidth]{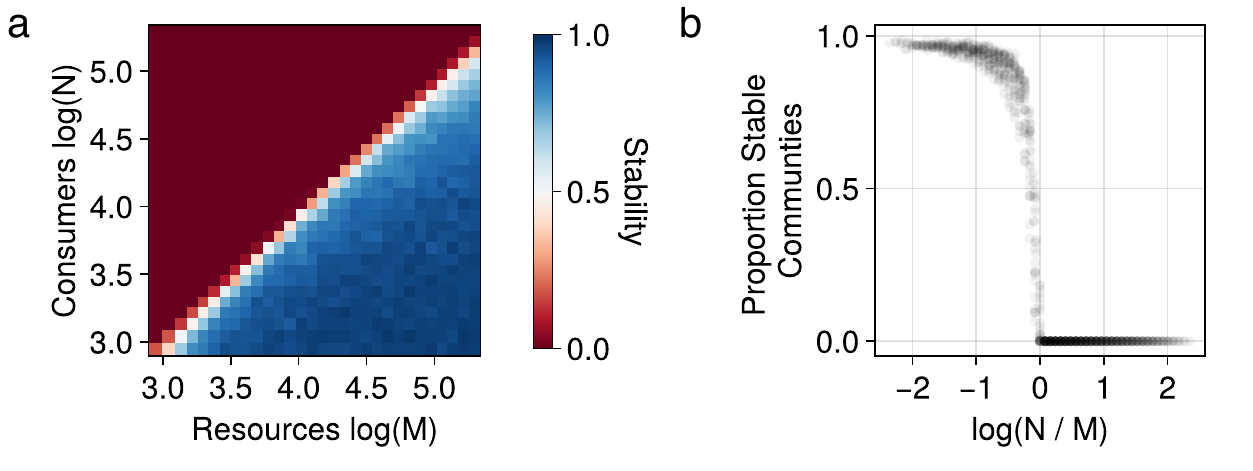}
    \caption{The effect of system size on microbial community stability. a) heatmap shows the average probability of stability across replicates as a function of the number of consumers $N$ (y-axis) and resources $M$ (x-axis) in log-scale. Stability is enhanced as the number of resources increases but declines with increasing numbers of consumers. b) Scatter plot shows stability as a function of the logarithm of the consumer-resource ratio. Stability quickly declines as it approaches $N = M$ after which no system is stable.}
    \label{fig:size_stability}
\end{figure}

Microbial community stability shows strong responses to system size, increasing with the number of resources $M$ but decreasing with the consumer richness $N$ (\cref{fig:size_stability}a). There is a strict stability boundary at $N=M$ beyond which systems with more consumers than resources ($N>M$) are universally unstable, recovering the classic species-packing limit \parencite{levin_community_1970}. The transition is most evident when plotting stability against the consumer-resource ratio (\cref{fig:size_stability}b), where communities display rapid destabilization as the ratio $N/M$ approaches $1$. Below the limit, when $M \geq N$, systems become more stable as the consumer-resource ratio falls. 

The observed instability when $N > M$ is a fundamental consequence of the dimensionality of the system and its ability to constrain dynamics around the equilibrium. For a fixed point to be stable, the resource environment must be able to provide enough independent ways to regulate every consumer population. Mathematically this requires the Jacobian matrix to be \textit{full rank} such that $\text{rank}(\mathbf{J}) = N+ M$. If the rank is deficient ($\text{rank}(\mathbf{J}) < N+ M$), the system will have at least one zero-valued eigenvalue which corresponds to a neutrally stable manifold along which the system can drift. We can express the source of this instability by first noting the block structure of the Jacobian
\begin{align}
    \mathbf{J} = \begin{bmatrix}
        \mathbf{J}_{cc} & \mathbf{J}_{cr} \\ \mathbf{J}_{rc} & \mathbf{J}_{rr}
    \end{bmatrix},
\end{align}
where $\mathbf{J}_{cc}$ and $\mathbf{J}_{rr}$ are the direct intra-compartment feedbacks within consumers and resources, and $\mathbf{J}_{cr}$ and $\mathbf{J}_{rc}$ are the coupling of consumers to resources and vice versa. In this block form, the total rank of the Jacobian is bounded by the rank of the rows
\begin{align}
    \text{rank}(J) \leq \text{rank}([\mathbf{J}_{cc} \ \mathbf{J}_{cr}]) + \text{rank}([\mathbf{J}_{rc} \ \mathbf{J}_{rr}]).
\end{align}
As we assume the self-limitation terms of consumers are the same $g^{(x)} = m^{(x)}$, the consumer block is equal to zero $\mathbf{J}_{cc} = 0$, allowing us to write the bound directly in terms of system size
\begin{align}
    \text{rank}(\mathbf{J}) \leq \text{min}(N,M) + M.
\end{align}
If the number of consumers is greater than the number of resources $N > M$, the rank is bounded at $\text{rank}(\mathbf{J}) \leq 2M < N + M$, which guarantees that the Jacobian is rank deficient and will be unstable with at least one zero valued eigenvalue.

\subsection*{Metabolic dynamics govern stability in feasible communities}
Having established the broad-scale patterns between system diversity and stability, we now explore the factors governing stability within the space of feasible system sizes ($N \leq M$). We consider how the key structural and dynamic parameters in the model influence stability across an ensemble of $10^5$ communities of fixed size ($N=75$,~$M=100$). The network topology and structural parameters of each community are generated as previously described, though elasticity parameters are sampled from broader ranges without the constraints on consumer self-limitation (see Table S1). 

For each realization, we assess community stability via the sign of the real part of the dominant eigenvalue. We characterize the impact of each parameter using the standard approach in generalized modeling, by calculating its correlation with stability across the entire ensemble \parencite{massing_generalized_2022}. Positive correlations identify stabilizing factors, while negative correlations highlight destabilizing factors. 

To gain a better understanding of the mechanistic action of these parameters, we also analyze the structure of the dominant dynamical mode across unstable communities. The least-stable mode captures the combination of consumers and resources that align with the dominant eigenvalue $\lambda_{\text{max}}$, dictating the trajectory of escape from the fixed point. We compute the proportion of this mode concentrated on the consumer component
\begin{align}
    B_c = \sum_{i=1}^N |v_i|^2,
\end{align}
where $||v_i||^2$ is the magnitude of the $i^{\text{th}}$ element of the dominant eigenvector. When $B_c = 1$ it indicates that all instability comes from consumers whereas $B_c = 0$ indicates instability comes from the resource fluctuations. We correlate this metric with each parameter value to characterize its mode of action, correcting the sign of the correlation based on the parameters effect on stability. A positive value indicates that the parameter acts through the consumer component and a negative value indicates it acts via the resource component.

\begin{figure}[H]
    \centering
    \includegraphics[width=\textwidth]{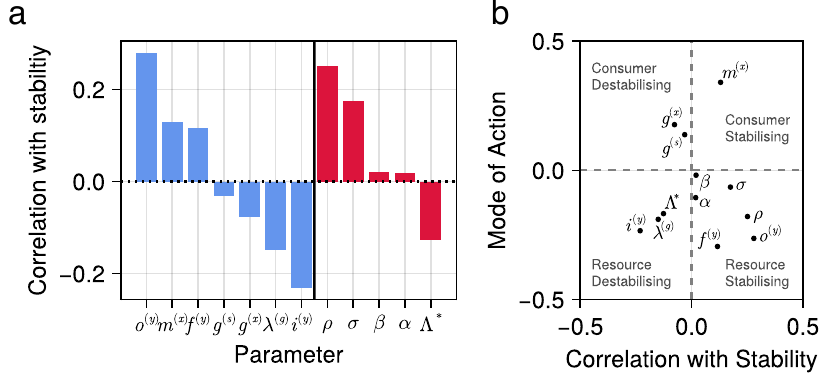}
    \caption{Correlation of stability with structural and dynamic parameters in microbial communities. a) Each bar shows the correlation of a given parameter (x-axis) with stability across all replicate communities. Bars are Coloured based on the type of parameter, blue for elasticity parameters and red for structural parameters. b) Scatter plot showing the correlation of each parameter with stability (x-axis) and the mode of action, measured with the sign corrected localization metric. Each point is a single parameter, as indicated by its corresponding label. The location indicates if the parameter is stabilizing (left to right) and which part of the system it acts on (top to bottom) as indicated by the labels in each quadrant.}
    \label{fig:dynamic_stability}
\end{figure}

\Cref{fig:dynamic_stability}a illustrates the effect of each parameter on community stability. The most pronounced effects stem from parameters controlling the self-feedback loops of individual populations or resources. Specifically, the elasticity of the self-limitation terms, mortality $m^{(x)}$ and outflow $o^{(y)}$, are strongly stabilizing while the terms for uptake $g^{(x)}$ and resource inflow $i^{(y)}$ are destabilizing. The mechanism for these shifts is revealed by the orientation of the least-stable dynamical mode (\cref{fig:dynamic_stability}b). The stabilizing parameters ($m^{(x)}$, $o^{(y)}$) act by suppressing growth of their respective compartments while the consumer gain and resource inflow ($g^{(x)}$, $i^{(y)}$), promote growth. Together, these results underscore the importance of self-limitation: when the net feedback of consumers or resources on themselves is positive, small perturbations will be amplified, creating instability. 

Next are the parameters that influence stability by controlling how fluctuations propagate between the consumer and resource parts of the system. The sensitivity of uptake to substrate abundance ($g^{(s)}$) is weakly negatively correlated with stability, tending to destabilize the consumer component. This reflects a trade-off in its role in community dynamics: while $g^{(s)}$ destabilizes the system by amplifying consumer growth, it simultaneously provides stability by strengthening the negative feedback on the resource compartment.

In contrast, the action of resource-switching plasticity, $f^{(y)}$, is much clearer, acting as a stabilizing force on resources. By allowing consumers to dynamically shift uptake towards the most abundant resources, $f^{(y)}$ promotes the dampening of resource fluctuations. Conversely, both the baseline consumer by-product release ($\Lambda^*$) and its sensitivity to uptake rates ($\lambda^{(g)}$) reduce stability by destabilizing the resource component. This occurs through a positive feedback loop: as efficiency drops, a larger fraction of uptake is released back into the environment, leading to an uncontrolled build-up of resources. This is exacerbated by $\lambda^{(g)}$, which causes by-product release to increase in response to resource abundance, amplifying the feedback loop.

Finally, the parameters that govern the relative importance of the environmental exchange, $\rho$ and $\sigma$, show a significant positive correlation with stability, acting to stabilize the resource compartment. As these values approach $1$, the system becomes increasingly dominated by external flux rather than internal cycling. This allows the system to buffer against perturbations: when internal dynamics create a spike in resource abundance, the excess is quickly flushed out, preventing the accumulation of resources that could otherwise cause instabilities. 

\subsection*{Trophic structure stabilizes cross-feeding communities}
Finally, we consider how the structure of consumer uptake and metabolic cross-feeding shapes community stability. Specifically, we focus on trophic structure, the hierarchical organization of communities across sequential degradation steps. Trophic structure has been suggested as a general organizing principle in microbial communities \parencite{gralka_trophic_2020}, and has been observed empirically, for example, in communities degrading complex marine polymers \parencite{datta_microbial_2016}.

\begin{figure}[H]
    \centering
    \includegraphics[width=1.0\textwidth]{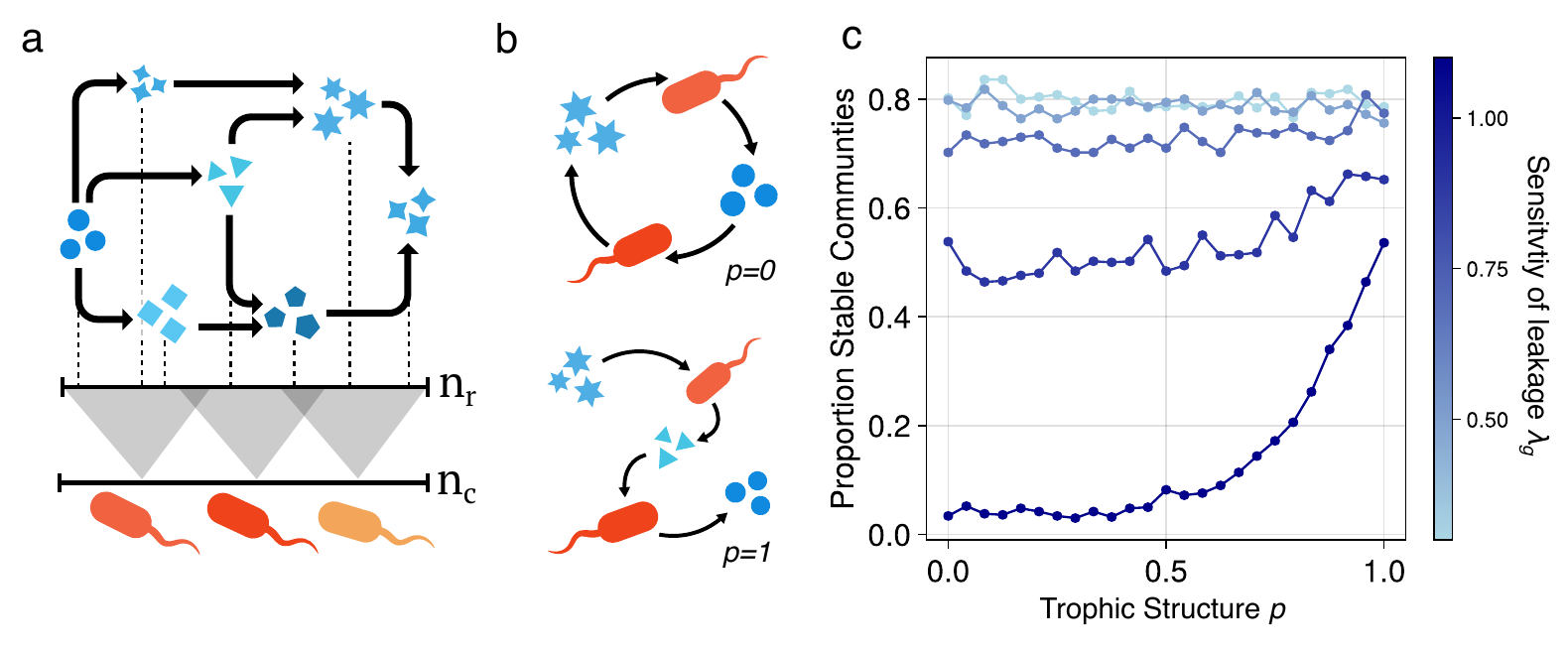}
    \caption{\textbf{Trophic structure stabilizes cross-feeding communities} a) Diagram showing the niche model used to generate trophic structure. Each consumer and resource is placed along a niche axis $n$. Consumers use resources within a range and the metabolism of resources creates products downstream, forming the hierarchical reaction network above. b) Motifs showing the structure of cross-feeding feedback loops present in random networks (top) and the directed degradation created by trophic structure (bottom). c) Plot showing how community stability (y-axis) varies with the degree of trophic structure $p$ (x-axis). Each line shows the proportion of stable communities across the ensemble with a given mean sensitivity of leakage $\lambda^{(g)}$. When communities have random structure ($p \approx 0$) they become unstable as the sensitivity of leakage increases. As trophic structure increases, the effect of leakage sensitivity becomes much weaker, allowing communities to remain more stable.}
    \label{fig:trophic}  
\end{figure}

To assess the impact of this trophic structure, we generate an ensemble of $1.25 \times 10^5$ communities across a continuum from hierarchical to random topologies. We construct trophic networks using an approach similar to the niche model used in food web ecology \parencite{williams_simple_2000}, assigning each consumer and resource a niche position, $n$, along a one-dimensional axis (\cref{fig:trophic}a). Consumers utilize resources located within a randomly drawn range around their niche value, while the metabolism of each resource is restricted to create by-products with higher values on the niche axis. This creates the hierarchical structure in which all by-product generation occurs in one direction, preventing the formation of loops. We control the degree of trophic structure by randomly rewiring interaction links with probability $1-p$, such that $p = 1$ yields maximal trophic structure and $p=0$ yields fully random communities. Branching and elasticity parameters are sampled from the same distributions as those used in the preceding section, though we fix the elasticity of consumer mortality $m^{(x)} = g^{(x)}$ to isolate the effect of network topology from consumer self-limitation.

We find that trophic structure stabilizes communities by buffering against destabilizing feedback loops created by metabolic cross-feeding. In unstructured communities ($p \approx 0$), cross-feeding creates cyclical pathways where metabolites act as indirect precursors to themselves (\cref{fig:trophic}b). When by-product leakage is insensitive to uptake ($\lambda^{(g)} \approx 0$), these loops have neutral effects on stability; however, when leakage accelerates with consumption ($\lambda^{(g)} > 0$), they trigger runaway instabilities (\cref{fig:trophic}c). Imposing hierarchical trophic structure ($p \to 1$) allows communities to remain stable even as the sensitivity of by-product leakage increases. This happens because trophic structure structurally breaks feedback loops, forcing by-products to flow downstream to be cleared by subsequent degradation or outflow. 

\section*{Discussion}
Understanding the stability of microbial communities presents a challenge due to the inherent complexity of their dynamics. By using the generalized modeling framework, we have developed and analyzed a model of microbial cross-feeding, bypassing the limitations of standard stability analysis. Our approach lets us identify how size, dynamics and structure influence stability in these systems. We find that microbial stability is strongly promoted by population self-limitation, metabolic plasticity, and hierarchical trophic structures. Conversely, stability is reduced by metabolic inefficiency, feedback loops and, most fundamentally, high consumer-resource ratios.

By exploring the limits on system size, our results clarify how cross-feeding facilitates the persistence of diverse microbial communities. Our findings show how these systems remain bound by the constraints of resource dimensionality, including the strict species-packing limit ($N \leq M$). This stands in contrast to both the extreme diversity of microbial communities seen in nature and experiments showing how even on simple substrates like glucose, diverse communities can persist due to by-product exchange \parencite{goldford_emergent_2018}. This apparent contradiction is resolved by considering cross-feeding not as a dynamical stabilizer, but rather one that acts via its structural effects. By expanding the effective dimensionality of resource space ($M$) via the creation of novel by-products, cross-feeding allows communities to escape the bounds on diversity even when few resources are supplied \parencite{marsland_available_2019}. This point is further supported by recent work demonstrating the massive configuration space for even relatively small organic molecules \parencite{shojaei_counting_2026}, suggesting how limits imposed by dimensionality of resource space may be relatively easy to avoid in microbial communities.

Within the space of feasible system sizes, our results show the dynamical features that promote stability. Beyond the expected stabilizing influence of self-limitation, our results highlight the ``openness'' ($\rho,\sigma$) of the microbial communities as a stabilizing feature: in the absence of other mechanisms systems dominated by external fluxes should tend to be more stable than closed systems driven by internal metabolite cycling. This finding makes sense; when systems are open they are effectively tied to the external environment allowing small fluctuations to quickly dissipate. 

In nature, however, many diverse microbial communities such as gut and soil microbiota operate near the closed end of this spectrum. Because our results and previous work on cross-feeding motifs suggest that strong metabolite cycling is inherently destabilizing \parencite{massing_generalized_2026}, the persistence of these closed communities implies a reliance on stabilizing mechanisms that counter these effects. As we show, one potential mechanism is the organization into hierarchical trophic structures, as seen in degrading communities like the gut \parencite{solden_interspecies_2018}. Beyond this, closed systems frequently rely on complementary stabilizing mechanisms, for example, forming biofilms which create spatial compartmentalization \parencite{nadell_sociobiology_2009} or evolving obligate metabolic interdependencies like auxotrophies \parencite{zelezniak_metabolic_2015}. Together, these features all act to dampen resource fluctuations and prevent runaway population growth that might otherwise lead to instability. 

A unique strength of our generalized approach is that it allows us to capture complex metabolic strategies without defining specific functional forms. We find that while metabolic plasticity ($f^{(y)}$) promotes stability, the inefficiency of metabolism and its response to the environment ($\Lambda^*,\lambda^{(y)}$) tend to destabilize communities. These results align with observations of community assembly trajectories, in which fast-growing but metabolically inefficient communities are replaced by more efficient configurations \parencite{cordero_microbial_2016}. Our findings map onto this process directly; the inherent instability of initial communities is replaced by configurations that have more efficient and thus stable dynamics. This difference in the stability of plasticity and inefficiency also highlights an interesting evolutionary conflict, as while engaging in inefficient overflow metabolism may maximize individual short-term fitness, it inherently leads to community instability, reducing long-term persistence. 

In this work we have focused primarily on \emph{local stability} of microbial communities by analyzing linearized dynamics around a fixed point. While a system is mathematically unstable if it departs from this equilibrium, it is important to note that this does not necessarily imply ecological collapse. On one hand, the instability of a fixed point may simply lead the community towards a new stable configuration, which may be structurally similar. Alternatively, community persistence can often be maintained through non-linear forces operating away from the fixed point, resulting in bounded dynamics such as limit cycles or chaotic attractors. This distinction is particularly relevant regarding the rank deficiency when $N > M$, which yields zero-valued eigenvalues corresponding to neutrally stable manifolds. Although these manifolds are unstable in a strictly local-linear sense they may be stabilized via higher-order mechanisms. While these non-linear properties could theoretically be integrated into our framework, it would require significantly more information to characterize the higher-order derivatives in the system.

\printbibliography
\end{document}


\maketitle

\section*{Model Derivation}
We start with a general model of microbial community dynamics which tracks the biomass of $N$ microbial consumer populations and the abundance of $M$ resources in the environment. 
\begin{align}
  \dot{X}_i &= G_i(X_i, \mathbf{Y}) - L_i(G_i) - M_i(X_i) \label{eq:dX} \\
  \dot{Y}_a &= I_a(Y_a) - O_a(Y_a) - \sum_i U_{ia}(X_i, \mathbf{Y}) + \sum_i K_{ia}(X_i, \mathbf{Y}). \label{eq:dY}
\end{align}

The change in consumer biomass over time $X_i$ is determined by the uptake of resources $G_i$ minus the loss of resources via the secretion of metabolites $L_i$ minus a mortality term $M_i$. Resources, $Y_a$, are supplied and removed externally at rates $I_a$ and $O_a$ respectively.  Resources are also taken up from the environment via consumer uptake at rate $U_{ia}$ which is the rate of loss of resource $a$ caused by consumer $i$ and added back to the environmental pool via by-product release $K_{ia}$ which is the amount of resource $a$ leaked by consumer $i$. All fluxes are expressed as general non-linear functions. The generalized modeling method does not require that we define their specific form, allowing us to consider dynamics across a range of potential functions. This aspect is discussed in more detail below. 

This model sets out the broad structure of the microbial community dynamics but does not explicitly include the linkages between consumers and resources which occur via uptake and by-product release. To include this we need to introduce additional constrains on the gain and loss terms.

\paragraph{By-product release}
We first consider the linkage between by-product release and uptake by consumers. We assume that metabolite production happens as a result of metabolic activity in the cell. This means that the rate of by-product release is proportional to the rate at which resources are taken up from the environment and used to fuel metabolism. We represent this constraint by defining a the variable $\Lambda_i(G_i)$ which gives the proportion of uptake that is released as by-products (i.e., the overall inefficiency of consumer metabolism)
\begin{align}
    \label{eq:Lambda}
    \Lambda_i(G_i) = \frac{L_i(G_i)}{G_i(X_i, \mathbf{Y})}.
\end{align}

To determine which resources are released as by-products we also define a stoichiometric matrix $D_{ab}$ which gives the proportion of by-products from the uptake of resource $a$ that are released as resource $b$. The value of $D_{ab}$ is constrained so that the proportions over all products sum to one 
\begin{align}
    \sum_b D_{ab} = 1.
\end{align}
$D_{ab}$ represents the whole set of metabolic pathways that consumers can use and is therefore shared across all populations the community. 

\paragraph{Uptake Rates} Next we consider how the total consumer uptake rates $G_i(X_i, \mathbf{Y})$ are linked to the gain and loss of resources from the environmental pool $U_{ia}(X_i, \mathbf{Y})$ and $K_{ia}(X_i, \mathbf{Y})$.

We start by considering the dependence of uptake rates on the substrates available to each consumer. We define the quantity $S_i(\mathbf{Y})$ to represent the total amount of substrates available for consumer $i$, accounting for both consumer preferences and the abundance of the resources in the environment. $S_i(\mathbf{Y})$ is defined as a sum across all resource contributions
\begin{align}
    S_i(\mathbf{Y}) &= \sum_a F_{ia}(Y_a), \label{eq:S}
\end{align}
where the functions $F_{ia}(Y_a)$ gives the contribution of resource $a$ to the substrates of consumer $i$. This lets us rewrite the total consumer uptake as a function of the new substrate abundance variable $S_i(\mathbf{Y})$
\begin{align}
    G_i(X_i, \mathbf{Y}) = G_i(X_i, S_i). \label{eq:G}
\end{align} 

The new quantities $S_i(\mathbf{Y})$ and $F_{ia}(Y_a)$ also let us link consumer uptake and resource gain and loss. Since resource $a$ contributes the fraction $F_{ia}(Y_a) / S_i(\mathbf{Y})$ to the total uptake of consumer $i$ we can express the loss from the environmental pool due to this uptake as
\begin{align}
    U_{ia}(X_i,\mathbf{Y}) &= \frac{F_{ia}(Y_a)}{S_i(\mathbf{Y})} G_i(X_i, S_i). \label{eq:U}
\end{align}
Likewise the growth of the resource pool due to by-product leakage can be expressed in terms of the proportion of total consumer uptake that is released as resource $a$ across all $b$ resources
\begin{align}
    K_{ia}(X_i,\mathbf{Y}) &=  \Lambda_i(G_i) \sum_b \frac{F_{ib}(Y_b)}{S_i(\mathbf{Y})} G_i(X_i, S_i) D_{ba}
\end{align}
where the inclusion of $\Lambda_i(G_i)$ and $D_{ba}$ account for the both the overall efficiency of the metabolism of the consumer and the stoichiometry of metabolism. 

\subsection*{Normalization}
Following the generalized modeling framework we next we normalize \cref{eq:dX,eq:dY} relative to steady state, using the lowercase notation to indicate normalized variables and $^*$ to indicate equilibrium quantities. We have omitted the function arguments for clarity. First the state variables
\begin{align}
    x_i &= \frac{X_i}{X_i^*} \label{eq:x_norm}\\
    y_a &= \frac{Y_a}{Y_a^*} \label{eq:y_norm}
\end{align}

and the functions:
\begin{align}
    g_i(x_i,s_i) &= \frac{G_i}{G_i^*} \label{eq:g_norm}\\
    l_i(g_i) &= \frac{L_i}{L_i^*} \label{eq:l_norm}\\
    m_i(x_i) &= \frac{M_i}{M_i^*} \label{eq:m_norm}\\
    i_a(y_a) &= \frac{I_a}{I_a^*} \label{eq:i_norm}\\
    o_a(y_a) &= \frac{O_a}{O_a^*} \label{eq:o_norm}\\
    u_{ia}(x_i,\mathbf{y}) &= \frac{U_{ia}}{U_{ia}^*} \label{eq:u_norm}\\
    k_{ia}(x_i,\mathbf{y}) &= \frac{K_{ia}}{K_{ia}^*} \label{eq:k_norm} \\
\end{align}

Next we consider how the auxiliary variables $\Lambda_i$ and $S_i$ are affected by normalization. The normalized proportion of by-product release $\Lambda_i$ from \cref{eq:Lambda} is given by
\begin{align}
    \lambda_i(g_i) &= \frac{\Lambda}{\Lambda^*} = \frac{L_i}{G_i}\frac{G_i^*}{L_i^*} = \frac{l_{i}(g_i)}{g_i(x_i, s_i)}, \label{eq:lambda_norm}
\end{align}

and the equation for the available substrate $S_i$ in \cref{eq:S} becomes
\begin{align}
    s_{i}(\mathbf{y}) &= \frac{S_i}{S_i^*} = \sum_a \frac{F_{ia}^* }{S_i^*} f_{ia}(y_a) = \sum_a \chi_{ia} f_{ia}(y_a), \label{eq:s_norm}
\end{align}
where we define $\chi_{ia} = F_{ia}^* / S_i^*$ as the relative contribution of resource $a$ to the uptake of consumer $i$ at equilibrium. 

We also consider the expanded normalized forms of the resource loss $u_{ia}$ 
\begin{align}
    u_{ia}(x_i, \mathbf{y}) = \frac{U_{ia}}{U^*_{ia}} = \frac{f_{ia}(y_a)}{s_i(\mathbf{y})} g_i(x_i,s_i)
\end{align}
and the leakage gain $k_{ia}(x_i,\mathbf{y})$
\begin{align}
    k_{ia}(x_i,\mathbf{y}) = \frac{K_{ia}}{K^*_{ia}} = \sum_b  \phi_{iba} \frac{f_{ib}(y_a)}{s_i(\mathbf{y})} g_i(x_i,s_i) \lambda_i(g_i)
\end{align}
where $\phi_{iba}$ is an additional parameter giving the proportion of uptake of $b$ by consumer $i$ released as $a$, defined in terms of the consumer uptake preferences and the stoichiometric matrix
\begin{align}
    \phi_{iba} = \frac{\chi_{ib} D_{ba}}{\sum_b \chi_{ib} D_{ba}}.
\end{align}
and is normalized across substrates such that $\sum_b \phi_{iba} = 1$.

Substituting the new normalized variables from \crefrange{eq:g_norm}{eq:s_norm} into the equation lets us rewrite the full normalized dynamic equations as:

\begin{align}
    \label{eq:norm_x}
    \frac{dx_i}{dt} &= \frac{G^*}{X^*} g_i(x_i,s_i) - \frac{G^* \Lambda^*}{X^*} g_i(x_i,s_i) \lambda(g_i)  - \frac{M^*}{X^*} m_i(x_i) \\ 
    \label{eq:norm_y}
    \frac{d y_a}{dt} &= \frac{I^*}{Y^*} i_a(y_a) - \frac{O^*}{Y^*} o_a(y_a) - \sum_i \frac{U^*}{Y^*} u_{ia}(x_i,\mathbf{y}) + \sum_i \frac{K^*}{Y^*} k_{ia}(x_i, \mathbf{y})
\end{align}

\subsection*{Scale and Branching Parameters}
Next we define the scale and branching parameters which determine the timescale of turnover and relative importance of individual fluxes for consumers and resources respectively

First, we compute the scale parameters $\alpha_i$ and $\beta_a$ for consumer $i$ and resource $a$ respectively by looking at the equilibrium of \cref{eq:norm_x,eq:norm_y}.
\begin{align}
    \alpha_i &= \frac{G^*}{X^*} = \frac{G^* \Lambda^*}{X^*} + \frac{M^*}{X^*} \label{eq:fixed_point_x}\\
    \beta_a &= \frac{I^*}{Y^*} + \sum_i \frac{K^*}{Y^*} = \frac{O^*}{Y^*} + \sum_i \frac{U^*}{Y^*} \label{eq:fixed_point_y}
\end{align}
These define the overall scale of fluxes of each consumer and resource at equilibrium. 
 
Second, we consider the branching parameters which define the relative size of fluxes in and out of the resource pools and are defined in terms of the scale parameters $\alpha_i$ and $\beta_a$. The proportion of resource gain that comes from external supply vs leakage is given by 
\begin{align}
    \rho_a &= \frac{1}{\beta_a}\frac{I^*}{Y^*} \\
    \tilde{\rho_a} &= 1 - \rho_a = \frac{1}{\beta_a}\sum_i \frac{K^*}{Y^*} 
\end{align}
where $\rho$ gives the proportion of resource gain from external supply whilst its inverse $\tilde{\rho}_a$ is the gain from leakage. We also define the equivalent branching parameters for the resource loss terms 
\begin{align}
    \sigma_a &= \frac{1}{\beta_a} \frac{O^*}{Y^*} \\
    \tilde{\sigma_a} &= 1 - \sigma_a = \frac{1}{\beta_a} \sum_i \frac{U^*}{Y^*}
\end{align}
where $\sigma_a$ is the loss of resources to the external environment and $\tilde{\sigma}_a$ is the loss due to uptake.

In addition to the overall balance of external and internal fluxes across the system we also define a set of branching parameters to represent the relative size of uptake and leakage fluxes between each resource and consumer pair. These parameters are given by     
\begin{align}
    \gamma_{ia} &= \frac{1}{\beta_a \tilde{\sigma}_a} \frac{U_{ia}^*}{Y^*} \\
    \eta_{ia} &= \frac{1}{\beta_a \tilde{\rho}_a} \frac{K_{ia}^*}{Y^*} 
\end{align}
where $\gamma_{ia}$ gives the contribution to total uptake of resource $a$ by consumer $i$ and $\eta_{ia}$ is the relative contribution of consumer $i$ to release of resource $a$.

Together these all let us write the final normalised equations
\begin{align}
    \frac{dx_i}{dt} &= \alpha_i \biggl[ g_i(x_i,s_i) - \Lambda_i^* g_i(x_i,s_i) \lambda_i(g_i) - (1- \Lambda_i^*) m_i(x_i) \biggr] \\
    \frac{dy_a}{dt} &= \beta_a \biggl[\rho_a i_a(y_a)
    - \sigma_a o_a(y_a) 
    - \tilde{\sigma}_a \sum_i \gamma_{ia} u_{ia}(x_i,\mathbf{y})
    + \tilde{\rho}_a \sum_i \eta_{ia} k_{ia}(x_i, \mathbf{y}) \biggr]
\end{align}
where 
\begin{align}
    u_{ia}(x_i, \mathbf{y}) &= \frac{f_{ia}}{s_i} g_i(x_i,s_i) \\
    k_{ia}(x_i,\mathbf{y}) &=  \sum_b  \phi_{iba} \frac{f_{ib}}{s_i} g_i(x_i,s_i) \lambda_i(g_i)
\end{align}

\subsection*{Elasticity Parameters}
With the normalized system in hand we now introduce the final set of parameters of the generalized model, the elasticities. These parameters are defined as the derivatives of the normalized functions at equilibrium and are equivalent to logarithmic derivatives (i.e $d\log(x)/d\log(y)$). They capture the non-linearity of processes central to the dynamics of the microbial community. For example, the saturation of consumer uptake is captured in the elasticity of uptake to substrate availability
\begin{align}
    g_i^{(s)} = \left. \frac{\delta g_i}{\delta s_i} \right\rvert_{*}
\end{align}
where the notation $g^{(s)}$ indicates the derivative of $g_i(x_i, s_i)$ with respect to $s_i$ evaluated at equilibrium. An elasticity parameter with value of 1 indicates that the function is linear, values above 1 indicate an elastic response where the function is more sensitive and vice versa. In this example the elasticity approaches a value of $0$ as consumers uptake becomes fully saturated.

\subsection*{Jacobian}
To compute the full Jacobian matrix we proceed as normal, linearising the system around the equilibrium point. As the normalized functions evaluate to 1 at equilibrium the resulting Jacobian is expressed purely in terms of the scale, branching and elasticity parameters we defined above. This means we don't have to explicitly solve for the steady state, avoiding one of the most common problems when considering the stability in non-linear systems.  

Applying this procedure to the consumer equations we get the first entries of the Jacobian
\begin{align}
    J_{ii} &= \alpha_i \left[g_i^{(x)} (1 - \Lambda^* (1 + \lambda_i^{(g)}) ) - (1 - \Lambda_i^*)m_i^{(x)} \right] \\
    J_{ij} &= 0 \quad \\
     J_{ia} &= \alpha_i g_i^{(s)} \chi_{ia} f_{ia}^{(y)} \left[1  -\Lambda_i^* (1 + \lambda_i^{(g)}) \right].
\end{align}
Here, the response of a consumer's growth to their own abundance is determined by the relative sensitivity of the uptake and mortality terms. The interspecific terms are zero as there is no direct dependence of consumer populations on one another. The dependence on resources is given by the gain from increased uptake, including the metabolic plasticity term $f_{ia}^{(y)}$ minus the effects of any change in overall efficiency. 

Next for the resource components we have
\begin{align}
    J_{ai} &= \beta_a g_i^{(x)} \left[
        \tilde{\rho}_a \eta_{ia}\left(1 + \lambda_i^{(g)}\right) - \tilde{\sigma}_a \gamma_{ia} 
    \right] \\
    J_{aa} &= \beta_a \left[ 
        \rho_a i_a^{(y)} - \sigma_a o_a^{(y)} - \tilde{\sigma}_a \sum_i \gamma_{ia} u_{ia}^{(y_a)} + \tilde{\rho}_a \sum_i \eta_{ia} k_{ia}^{(y_a)} 
        \right] \\
    J_{ab} &= \beta_a \left[ 
        -\tilde{\sigma}_a \sum_i \gamma_{ia} u_{ia}^{(y_b)} + \tilde{\rho}_a \sum_i \eta_{ia} k_{ia}^{(y_b)}  \right]
\end{align}

where

\begin{align}
    u_{ia}^{(y_b)} &= \delta_{ab} f_{ia}^{(y)} + (g_i^{(s)} - 1) \chi_{ib} f_{ib}^{(y)} \\
    k_{ia}^{(y_b)}  &= \phi_{iba} f_{ib}^{(y)} - \chi_{ib} f_{ib}^{(y)} [g_i^{(s)}(1 + \lambda_i^{(g)}) - 1]
\end{align}
and $\delta_{ab}$ is the Kronecker delta function. The response of resources to consumer abundance is determined purely by the change in consumer uptake, weighted by the amount of resource uptake and by-product release by that consumer. The diagonal terms include any changes to external supply and outflow minus resource uptake plus any by-product release. The uptake loss term is weighted for each consumer and is made up of the change in contribution of resource $y_a$ plus the change in total resource uptake by each consumer. The gain from by-products is determined by the by-product made by uptake across all $b$ resources in the system. The off-diagonal elements are very similar, but do not include the external supply, outflow and direct change in contributions to uptake and by-product release. 

\section*{Community sampling}
To generate the parameter set for a single community, we start with the adjacency matrices for uptake $A_U$ and stoichiometry $A_D$ as well as mathbftors of equilibrium growth $G^*$, leakage $\Lambda^*$ and scale parameters $\alpha$ and $\beta$. From these we need to derive the remaining branching parameters while ensuring the parameter set corresponds to a feasible fixed point, as determined by setting \cref{eq:norm_x,eq:norm_y} to zero:
\begin{align}
    0 &= \frac{G^*}{X^*} - \frac{G^*\Lambda^*}{X^*} - \frac{M^*}{X^*}\\
    0 &= \frac{I^*}{Y^*} - \frac{O^*}{Y^*} - \sum \frac{U^*}{Y^*} + \sum \frac{K^*}{Y^*} 
\end{align}

We first normalize the adjacency matrices to get the relative uptake matrix $\chi$ and stoichiometric matrix $D$
\begin{align}
    \chi_{ia} = \frac{A_{Uia}}{\sum_a A_{Uia}} \\
    D_{ab} = \frac{A_{Dab}}{\sum_b A_{Dab}}
\end{align}
We then use these normalized matrices to compute the total demand and supply for each resource by consumers
\begin{align}
    U^*_{tot} &= \chi^T G \\
    K^*_{tot} &= (\chi D)^T (G \cdot \Lambda)
\end{align}
where $(G\cdot\Lambda)$ is the (element-wise) product of consumer growth and inefficiency. These two terms correspond to the summation terms in the fixed point condition; thus, to ensure steady state, we can simply set the in- and outflow terms to balance the equation
\begin{align}
    I^*-O^* = U^*_{tot} + K^*_{tot}
\end{align}
Any resource that is demanded more than it is produced in the community will require extra input from the environment ($I*>0$), and and that is overproduced will require extra outflow to satisfy the fixed point condition ($O^* > 0$).

Having defined the external resource terms we can compute the relevant branching parameters 
\begin{align}
    \rho = \frac{I^*}{I^* + K^*_{tot}} \quad \quad \tilde{\rho} =1-\rho \\
    \sigma = \frac{O^*}{O^* + U^*_{tot}} \quad \quad \tilde{\sigma} =1-\sigma
\end{align}
which represent the balance of input and outflow from the resource pool between the external environment and the community.

We then compute the branching parameters that capture the structure of the consumer-resource interaction network. First we compute the total uptake and leakage flux matrices
\begin{align}
    U_{ia} &= G_i \chi_{ia} \\
    K_{ia} &= (G_i \Lambda_i) (\chi D)_{ia}
\end{align}
where $(\chi D)_{ia}$ is an element of the product of normalized uptake and stoichiometric matrix. From these we can compute the branching terms for uptake $\gamma$ and by-product release $\eta$ as
\begin{align}
    \gamma_{ia} = \frac{U^{\text{flux}}_{ia}}{\sum_i U^{\text{flux}}_{ia}} \\
    \eta_{ia} = \frac{K^{\text{flux}}_{ia}}{\sum_i K^{\text{flux}}_{ia}}
\end{align}
Finally we compute the uptake-leakage branching term
\begin{align}
    \phi_{iba} = \frac{\chi_{ib} D_{ba}}{\sum_b \chi_{ib} D_{ba}}
\end{align}
Together with the elasticity terms, these parameters let us compute the Jacobian matrix, ensuring it corresponds to a feasible fixed point.

\section*{Microbial Niche Model}
The microbial niche model generates the structure of uptake and leakage networks in a community with $N$ consumers and $M$ resources. To begin each consumer and resource is assigned a value along the niche axis $n$ drawn from a random distribution
\begin{align}
    \begin{split}
        n_i &\sim \text{Uniform}(0,1) \\ 
        n_a &\sim \text{Uniform}(0,1)
    \end{split}
\end{align}
We generate the structure of the uptake matrix $A_U$ by allowing each consumer to utilize resources that have niche values within a range with width $w_i$ centered around their own position $n_i$. We assign the width values by drawing from a beta distribution, parameterised so that the consumption network has the desired connectance $C$
\begin{align}
    2w_i \sim \text{Beta}\left(1, \frac{1}{C}-1 \right)
\end{align}
Consumers utilize resources that sit within this range. The uptake of resource $a$ by consumer $i$ is thus given by
\begin{align}
    I_{ia} = \begin{cases} 1 \quad \text{if} \quad n_i + w_i< n_a < n_i - w_i \\
    0 \quad \text{else}
    \end{cases}
\end{align}

The stoichiometric matrix is constructed in a similar manner, with each resource producing by-products that occur within a range with width $w_a$ centered around a point $c_a$ which is offset from the focal niche position, forcing the hierarchical structure of metabolism. We draw the width of the range from the same distribution to ensure connectance $C$
\begin{align}
    2w_a \sim \text{Beta}\left(1, \frac{1}{C}-1 \right)
\end{align}
We draw the center points relative position $p_a$ from a separate distribution
\begin{align}
    p_a \sim \text{Beta}\left(1, \frac{1}{d}-1 \right)
\end{align}
where $d$ is the mean relative distance to the end of the niche axis from the focal resource niche value $n_a$. We then use this to get the realized center point of by-product creation
\begin{align}
    c_a = n_a + (1 - n_a) p_a
\end{align}
We then construct the structure of the stoichiometric matrix $A_D$ such that each resource $a$ creates by product $b$ if it sits within it niche range
\begin{align}
    D_{ab} = \begin{cases} 1 \quad \text{if} \quad \text{max}(c_a + w_a, n_a) < n_b < c_a - w_a \\
    0 \quad \text{else}
    \end{cases}
\end{align}

\newpage

\begin{table}[h!]
    \caption{Parameter values for different community ensembles}
    \renewcommand{\arraystretch}{1.2}
    \begin{tabularx}{1.1\textwidth}{l|XXX}
    \hline
    \textbf{Parameter} & \textbf{Size} & \textbf{Dynamics} & \textbf{Trophic~Structure} \\ \hline
    $g^{(x)}$ & 0--1 & 0.0--0.5 & 0--1 \\
    $g^{(s)}$ & 0--2 & 0--2 & 0--2 \\
    $m^{(x)}$ & $g^{(x)}$ & 0.5--1.0 & $g^{(x)}$ \\
    $f^{(y)}$ & 1--2 & 0--2 & 0--2 \\
    $\lambda^{(g)}$ & 0 & 0--2 & -- \\
    $I^{(y)}$ & 0 & 0.0--0.5 & 0.0--0.5 \\
    $o^{(y)}$ & 0.5--1.5 & 0.5--1.0 & 0.5--1.0 \\
    $\Lambda$ & 0.0--0.9 & 0.0--0.9 & 0.5 \\
    $G$ & 0--1 & 0--1 & 0--1 \\
    $\alpha$ & 0--1 & 0--1 & 0--1 \\
    $\beta$ & 0--1 & 0--1 & 0--1 \\ \hline
    \end{tabularx}\label{tab:model_params}
\end{table}